\documentclass{article}
\usepackage{spconf,amsmath,graphicx}
\usepackage{booktabs}
\usepackage{caption}
\usepackage{subcaption}
\usepackage{dsfont}
\usepackage{bm}
\usepackage{setspace}
\usepackage{hyperref}

\title{A Variational Bayesian Approach to Learning Latent Variables for Acoustic Knowledge Transfer}
\name{Hu Hu$^{1}$,
      Sabato Marco Siniscalchi$^{1,2}$,
      Chao-Han Huck Yang$^{1}$,
      Chin-Hui Lee$^{1}$
      }
\address{$^1$School of Electrical and Computer Engineering, Georgia Institute of Technology, GA, USA \\
$^2$Computer Engineering School, University of Enna Kore, Italy\\
}
\begin{document}
%
\maketitle
\begin{abstract}
We propose a variational Bayesian (VB) approach to learning distributions of latent variables in deep neural network (DNN) models for cross-domain knowledge transfer, to address acoustic mismatches between training and testing conditions. Instead of carrying out point estimation in conventional maximum a posteriori estimation with a risk of having a curse of dimensionality in estimating a huge number of model parameters, we focus our attention on estimating a manageable number of latent variables of DNNs via a VB inference framework. To accomplish model transfer, knowledge learnt from a source domain is encoded in prior distributions of latent variables and optimally combined, in a Bayesian sense, with a small set of adaptation data from a target domain to approximate the corresponding posterior distributions. Experimental results on device adaptation in acoustic scene classification show that our proposed VB approach can obtain good improvements on target devices, and consistently outperforms 13 state-of-the-art knowledge transfer algorithms.

\end{abstract}

\vspace{-0.2cm}
\section{Introduction}
\label{sec:intro}
\vspace{-0.1cm}
Recent advances in machine learning are largely due to an evolution of deep learning combined with an availability of massive amounts of data. Deep neural networks (DNNs) have demonstrated state-of-the-art results in building acoustic systems \cite{hinton2012deep, seide2011conversational, xu2014regression}.
Nonetheless, audio and speech systems still highly depend on how close the training data used in model building covers the statistical variation of the signals in testing environments. Acoustic mismatches, such as changes in speakers and recording devices, usually cause an unexpected and severe performance degradation \cite{lee2000adaptive, li2014overview, bell2020adaptation}.
For example, as for acoustic scene classification (ASC), device mismatch is an inevitable problem in real production scenarios \cite{mesaros2018multi, gharib2018unsupervised, koutini2020low, hu2020relational}.
Moreover, the amount of data for the specific target domain is often not sufficient to train a good deep target model to achieve a similar performance to the source model. A key issue is to design an effective adaptation procedure to transfer knowledge from the source to target domains, while avoiding catastrophic forgetting and curse of dimensionality \cite{goodfellow2013empirical, huang2015maximum, kirkpatrick2017overcoming, powell2007approximate, poggio2017and} often encountered in deep learning.

\begin{figure*}[t]
  \centering
  \includegraphics[width=0.85\linewidth]{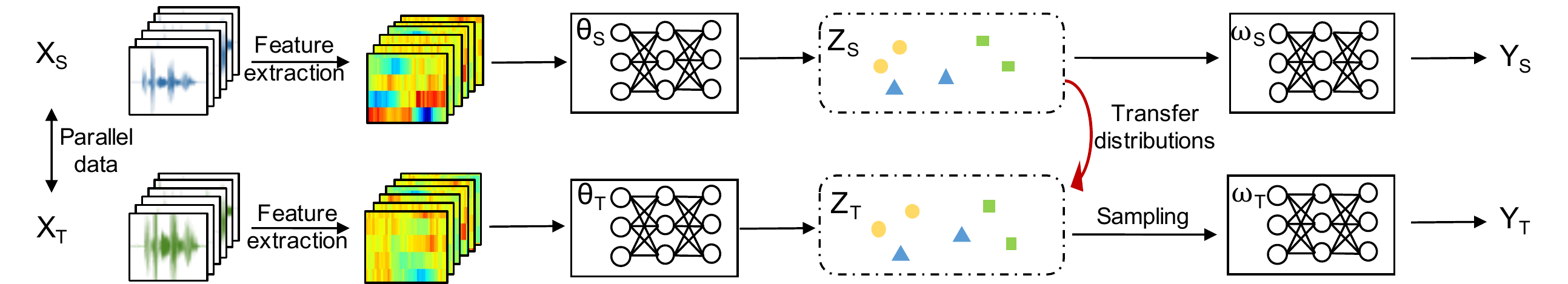}
  \caption{Illustration of the proposed knowledge transfer framework.}
  \label{fig:framework}
  \vspace{-0.4cm}
\end{figure*}

Bayesian learning provides a mathematical framework to model uncertainties and incorporate prior knowledge. It usually performs estimation via either maximum a posteriori (MAP) or variational Bayesian (VB) approaches. By leveraging upon target data and prior belief, a posterior belief can be obtained by optimally combining them. In the MAP solution, a point estimate can be obtained, which has been proven effective in handling acoustic mismatches in hidden Markov models (HMMs) \cite{lee2000adaptive, gauvain1994maximum, siohan2001joint} and DNNs \cite{huang2015maximum, huang2017bayesian, kirkpatrick2017overcoming} by assuming a distribution on the model parameters.
On the other hand, the VB approach performs an estimation on the entire posterior distribution via a stochastic variational inference method \cite{watanabe2004variational, graves2011practical, kingma2013auto, nguyen2017variational, hsu2018scalable, si2021variational}.
Bayesian learning can facilitate building an adaptive system for specific target conditions in a particular environment. Thus the mismatches between training and testing can be reduced, and the overall system performance is greatly enhanced.

Traditional Bayesian formulations usually impose uncertainties on model parameters, like Bayesian neural networks \cite{jospin2020hands}. However, for commonly used DNNs, the number of parameters is usually much larger than the available training samples, making an accurate estimation difficult. Moreover, a feature based knowledge transfer framework, namely teacher-student learning (TSL, also called knowledge distillation) \cite{li2014learning, hinton2015distilling} has been investigated in recent years. The basic TSL transfers knowledge acquired by the source / teacher model and encoded in its softened outputs (model outputs after softmax), to the target / student model, where the target model directly mimics the final prediction of the source model through a KL divergence loss.
The idea is then extended to hidden embedding of intermediate layers \cite{romero2014fitnets, zagoruyko2016paying, heo2019comprehensive}, where different embedded representations are proposed to encode and transfer knowledge. However, instead of considering the whole distribution of latent variables, they only perform point estimation as in MAP, potentially leading to sub-optimal results and may lose distributional information.

In this work, we aim at establishing a Bayesian adaptation framework based on latent variables of DNNs, where the knowledge is transferred in the form of distributions of deep latent variables.
Thus, a novel variational Bayesian knowledge transfer (VBKT) approach is proposed.
We take into account of the model uncertainties and perform distribution estimation on latent variables.
In particular, by leveraging upon variational inference, the distributions of the source latent variables (prior) are combined with the knowledge learned from target data (likelihood) to yield the distributions of the target latent variables (posterior).
Prior knowledge from the source domain is thus encoded and transferred to the target domain, by approximating the posterior distributions of latent variables.
An extensive and thorough experimental comparison against 13 recent cut-edging knowledge transfer methods is carried out. Experimental evidence demonstrates that our proposed VBKT approach outperforms all competing algorithms on device adaptation tasks of ASC.


\vspace{-0.2cm}
\section{Bayesian Inference of Latent Variables}
\label{sec:method}

\vspace{-0.2cm}
\subsection{Knowledge Transfer of Latent Variables}
\vspace{-0.1cm}
Suppose we are given some data observations $\mathcal{D}$, and let $\mathcal{D}_S = \{x_S^{(i)}, y_S^{(i)}\}^{N_S}_{i=1}$ and $\mathcal{D}_T = \{x_T^{(i)}, y_T^{(i)}\}^{N_T}_{i=1}$ indicate the source and target domain data, respectively.
Our framework requires parallel data, e.g., for each target data sample $x_T^{(i)}$, there exists a paired data sample $x_S^{(j)}$ from the source data, where $x_T^{(i)}$ and $x_S^{(j)}$ share the same audio content but recorded by different devices.
Consider a DNN based discriminative model with parameters $\lambda$ to be estimated, where $\lambda$ usually represents network weights. Starting from the classical Bayesian approach, a prior distribution $p(\lambda)$ is defined over $\lambda$, and the posterior distribution after seeing the observations $\mathcal{D}$ can be obtained by the Bayes Rule as follows,
\begin{equation}
    p(\lambda|\mathcal{D}) = \frac{p(\mathcal{D}|\lambda) p(\lambda)}{p(\mathcal{D})}.
    \label{eq:bayes}
\end{equation}

Figure~\ref{fig:framework} illustrates the overall framework of our proposed knowledge transfer approach. Parallel input features of $X_S$ and $X_T$ are firstly extracted and then fed into neural networks. In addition to network weights, we introduce the latent variables $Z$ to model the intermediate hidden embedding of DNNs. Here $Z$ refers to the unobserved intermediate representations, encoding transferable distributional information. We then decouple the network weights into two independent subsets, $\theta$ and $\omega$, as illustrated by the subnets in the 4 squares in Figure~\ref{fig:framework}, to represent weights before and after $Z$ is generated, respectively. Thus we have 
\begin{equation}
    p(\lambda) = p(Z, \theta, \omega) = p(Z|\theta) p(\theta) p(\omega).
    \label{eq:lambda}
\end{equation}

Note that the relationship in Eq.~(\ref{eq:lambda}) holds for both prior $p(\lambda)$ and posterior $p(\lambda|\mathcal{D})$. Here we focus on transferring knowledge in a distribution sense via the latent variables $Z$. With parallel data, we thus assume that there exists $Z$ retaining the same distributions across the source and target domains. Specifically, for the target model we have $p(Z_T|\theta_T) = p(Z_S|\theta_S, \mathcal{D}_S)$, as the prior knowledge learnt from the source encoded in $p(Z_S|\theta_S, \mathcal{D}_S)$.

\vspace{-0.2cm}
\subsection{Variational Bayesian Knowledge Transfer}
\vspace{-0.1cm}
Denoting $\lambda$ in the target model as $\lambda_T$, typically, the posterior $p(\lambda_T|\mathcal{D}_T)$ is often intractable, and an approximation is required. In this work, we propose a variational Bayesian approach to approximate the posterior; therefore, a variational distribution $q(\lambda_T|\mathcal{D}_T)$ is introduced. For the target domain model, the optimal $q^*(\lambda_T|\mathcal{D}_T)$ is obtained by minimizing KL divergence between the variational distribution and the real one, over a family of allowed approximate distributions $\mathcal{Q}$:
\begin{align}
    q^*(\lambda_T|\mathcal{D}_T) &= \mathop{argmin}\limits_{q\in \mathcal{Q}} \mathtt{KL}(q(\lambda_T|\mathcal{D}_T) \ \|\  p(\lambda_T|\mathcal{D}_T)).
    \label{eq:approx}
\end{align}

In this work, we focus on latent variables, $Z$, and we assume a non-informative prior over $\theta_T$ and $\omega_T$. Next, by substituting Eqs.~(\ref{eq:bayes}), (\ref{eq:lambda}), and the prior distribution into Eq.~(\ref{eq:approx}), we arrive, after re-arranging the terms, to the following variational lower bound $\mathcal{L}(\lambda_T; \mathcal{D}_T)$ as
\begin{align}
    \mathcal{L}(\lambda_T; \mathcal{D}_T) =\ & \mathds{E}_{Z_T \sim q(Z_T| \theta_T, \mathcal{D}_T)} \log p(\mathcal{D}_T |Z_T, \theta_T, \omega_T)\nonumber \\
    & - \mathtt{KL} (q(Z_T|\theta_T, \mathcal{D}_T)\ \|\ p(Z_S|\theta_S, \mathcal{D}_S)).
    \label{eq:elbo}
\end{align}

Simply put, a Gaussian mean-field approximation is used to specify the distribution forms for both the prior and posterior over $Z$. Specifically, each latent variable $z$ in $Z$ follows an $M$-dimension isotropic Gaussian,  where $M$ is the hidden embedding size.
Given a parallel data set, we can  approximate the KL divergence term in Eq.~(\ref{eq:elbo}) by establishing a mapping of each pair of Gaussian distributions across domains via sample pairs. We denote the Gaussian mean and variance for the source and target domains as $\mu_S; \sigma_S^2$ and $\mu_T; \sigma_T^2$, respectively. A stochastic gradient variational Bayesian (SGVB) estimator \cite{kingma2013auto} is then used to approximate the posterior, with the network hidden outputs being regarded as the mean of the Gaussian. Moreover, we assign a fixed value $\sigma^2$ to both $\sigma_S^2$ and $\sigma_T^2$, as the variance for all individual Gaussian components. We can now obtain a close-form solution for the KLD term in Eq.~(\ref{eq:elbo}). Furthermore, by adopting Monte Carlo to generate $N_T$-pairs of sample, the lower bound in Eq.~(\ref{eq:elbo}) can be approximated empirically as:
\begin{align}
    \mathcal{L}(\lambda_T; \mathcal{D}_T) =\ & \sum^{N_T}_i \mathds{E}_{z^{(i)}_T \sim \mathcal{N}(\mu_T^{(i)}, \sigma^2)} \log p(y_T^{(i)} | x_T^{(i)}, z^{(i)}_T, \theta_T, \omega_T)\nonumber \\
    & - \frac{1}{2\sigma^2} \sum_i^{N_T} \| \mu_T^{(i)} - \mu_S^{(i)}\|_2^2,
    \label{eq:elbo_new}
    \vspace{-0.2cm}
\end{align}
where the first term is the likelihood, and the second term is deduced from the KL divergence between prior and posterior of the latent variables. Each instance of $z^{(i)}_T$, is sampled from the posterior distribution as $z_T^{(i)}|\theta_T, \mathcal{D}_T \sim \mathcal{N}(\mu_T^{(i)}, \sigma^2)$, thus the expectation form in the first term can be reduced.
To flow the gradients of sampling operation through deep neural nets, a reparameterization trick \cite{salimans2013fixed, kingma2013auto} is adopted during the network training. In the inference stage, as it's a classification task, we directly take $z^{(i)}_T = \mu^{(i)}_T$ to simplify the computation.

\vspace{-0.2cm}
\section{Experiments}
\label{sec:exp}

\vspace{-0.2cm}
\subsection{Experimental Setup}
\vspace{-0.1cm}
We evaluate our proposed VBKT approach on the acoustic scene classification (ASC) task of DCASE2020 challenge task1a \cite{heittola2020acoustic}. The training set contains $\sim$10K scene audio clips recorded by the source device (device A), and 750 clips for each of the 8 target devices (Device B, C, s1-s6). Each target audio is paired with a source audio, and the only difference between the two audios is the recording device. The goal is to solve the device mismatch issue for one specific target device at a time, i.e., device adaptation, which is a common scenario in real applications. For each audio clip, log-mel filter bank (LMFB) features are extracted, and scaled to [0,1] before feeding into the classifier.

Two state-of-the-art models, namely: a dual-path resnet (RESNET) and a fully convolutional neural network with channel attention (FCNN), are tested according to the challenge results \cite{heittola2020acoustic, hu2020device}. We use the same models for both the source and target devices. Mix-up \cite{zhang2017mixup} and SpecAugment \cite{park2019specaugment} are used in the training stage. Stochastic gradient descent (SGD) with a cosine-decay restart learning rate scheduler is used to train all models. Maximum and minimum learning rates are 0.1, and 1e-5, respectively. The latent variables are based on the hidden outputs before the last layer. Specifically, the hidden embedding after batch-normalization but before ReLU activation of the second last convolutional layer is utilized. As stated in Section~\ref{sec:method}, a deterministic value is set for $\sigma$. In our experiments, we generate extra data \cite{hu2021two} and compute the average standard deviation over each audio clip, where we finally set $\sigma=0.2$. For the other 13 tested cut-edging TSL based methods, we mostly follow the recommended setups and hyper-parameter settings in their original papers. The temperature parameter is set to 1.0 for all when computing KL divergence with soft labels. \footnote{\footnotesize Code available: \url{https://github.com/MihawkHu/ASC_Knowledge_Transfer}}

\begin{table}[t]
\footnotesize
\centering
\caption{Comparison of average evaluation accuracies (in \%) on recordings of the DCASE2020 ASC data set. Each method is tested with and without the combination of the basic TSL method. Each cell represents the average value over 32 experimental results for 8 target devices $\times$ 4 repeated trails.}
\label{tab:res-all}
\vspace{-0.2cm}
\begin{tabular}{l||c|c|c|c}
\toprule
\toprule
Method      & \begin{tabular}[c]{@{}c@{}}RESNET\\ avg. (\%)\end{tabular} & \begin{tabular}[c]{@{}c@{}}RESNET \\w/ TSL\\ avg. (\%)\end{tabular} & \begin{tabular}[c]{@{}c@{}}FCNN\\ avg. (\%)\end{tabular} & \begin{tabular}[c]{@{}c@{}}FCNN \\w/ TSL\\ avg. (\%)\end{tabular} \\
\midrule
\midrule
Source model & 37.70 & - & 37.13 & - \\
No transfer & 54.29                                               & -                                                                       & 49.97                                              & -                                                                     \\
One-hot     & 63.76                                               & -                                                                       & 64.45                                             & -                                                                     \\
\midrule
TSL  \cite{hinton2015distilling}       & 68.04                                                 & 68.04                                                       & 66.27                                                & 66.27                                                      \\
NLE  \cite{meng2020vector}      & 65.64                                               & 67.76                                                        & 64.47                                             & 64.53                                                        \\
Fitnet \cite{romero2014fitnets}     & 66.73                                               & 69.89                                                        & 67.29                                             & 69.06                                                      \\
AT    \cite{zagoruyko2016paying}      & 63.73                                               & 68.06                                                        & 64.16                                             & 66.35                                                      \\
AB  \cite{heo2019knowledge}        & 65.34                                               & 68.69                                                        & 66.21                                             & 66.91                                                      \\
VID  \cite{ahn2019variational}       & 63.90                                               & 68.56                                                        & 63.79                                             & 65.75                                                      \\
FSP   \cite{yim2017gift}      & 64.44                                               & 68.94                                                        & 65.33                                             & 66.01                                                      \\
COFD   \cite{heo2019comprehensive}     & 64.92                                               & 68.57                                                        & 66.69                                             & 68.63                                                      \\
SP    \cite{tung2019similarity}      & 64.57                                               & 68.45                                                        & 65.74                                             & 67.36                                                      \\
CCKD  \cite{peng2019correlation}      & 65.59                                               & 69.47                                                        & 66.52                                             & 68.29                                                      \\
PKT   \cite{passalis2018learning}     & 64.65                                               & 65.43                                                        & 64.84                                             & 67.25                                                      \\
NST  \cite{huang2017like}      & 68.35                                               & 68.51                                                        & 67.13                                             & 68.84                                                      \\
RKD  \cite{park2019relational}      & 65.28                                               & 68.46                                                        & 65.63                                             & 67.27                                                      \\
\midrule
VBKT        & \textbf{69.58}                                               & \textbf{69.90}                                                        & \textbf{69.96}                                             & \textbf{70.50}                                                     \\
\bottomrule
\bottomrule
\end{tabular}
\vspace{-0.4cm}
\end{table}

\vspace{-0.2cm}
\subsection{Evaluation Results on Device Adaptation}
\vspace{-0.1cm}
Evaluation results of device adaptation on the DCASE2020 ASC task are shown in Table~\ref{tab:res-all}. The source models are trained on data recorded by Device A, where we can get a classification accuracy of 79.09\% for RESNET and 79.70\% for FCNN, on the source test set, respectively. There are 8 target devices, i.e., Device B, C, s1-s6. The accuracy reported in each cell of Table~\ref{tab:res-all} is obtained by averaging among 32 experimental results, from 8 target devices and 4 trails for each. The first and third columns represent results by directly using the knowledge transfer methods; whereas the second and fourth columns list accuracies obtained when further combined with the basic TSL method.

We look at the results without the combination of TSL at first. The 1st row gives results by directly testing the source model on target devices. We can observe a huge degradation (from $\sim$79\% to $\sim$37\%) when compared with the results on source test set. That shows the device mismatch is indeed a critical aspect in acoustic scene classification, as the device changing causing a serve performance drop. The 2nd and 3rd rows in Table~\ref{tab:res-all} give results of target model trained by target data either from the scratch or fine-tuned on source model. By comparing them we can argue the importance of knowledge transfer when building a target model.

The 4th to 16th rows in Table~\ref{tab:res-all} show the evaluated results of 13 recent top TSL based methods. The result of basic TSL \cite{li2014learning, hinton2015distilling}, which minimizes KL divergence between model outputs and soft labels, is shown in the 4th row. We can observe a gain obtained by TSL when compared with one-hot fine-tuning. If we compare the other methods (5th to 16th rows) with basic TSL, they only show small advantages on this task. The bottom row of Table~\ref{tab:res-all} shows results of our proposed VBKT approach. It not only outperforms one-hot fine-tuning by a large margin (69.58\% vs. 63.76\% for RESNET, and 69.12\% vs. 61.54\% for FCNN), but also attains superior classification results to those obtained with other algorithms.

We further investigate the combination of the proposed approach and the basic TSL. The experimental results are shown in the 2nd and 4th columns of Table~\ref{tab:res-all}, for RESNET and FCNN models, respectively. Specifically, the original cross entropy (CE) loss is replaced by an addition of 0.9 $\times$ KL loss with soft labels and 0.1 $\times$ CE loss with hard labels. There is no change to basic TSL so results remain the same in the 4th row. For the other tests (from 5th to 16th rows), when combining with basic TSL, most tested methods can attain further gains. Indeed, such a combination is recommended by some studies \cite{romero2014fitnets, zagoruyko2016paying, passalis2018learning}. Finally, we compare our proposed VBKT approach with others, when combined with TSL, the accuracy can be further boosted, and it still outperforms other tested methods under the same setup.

\begin{figure}[t]
     \centering
     \vspace{-0.2cm}
     \begin{subfigure}[b]{0.25\textwidth}
         \centering
         \includegraphics[width=\textwidth]{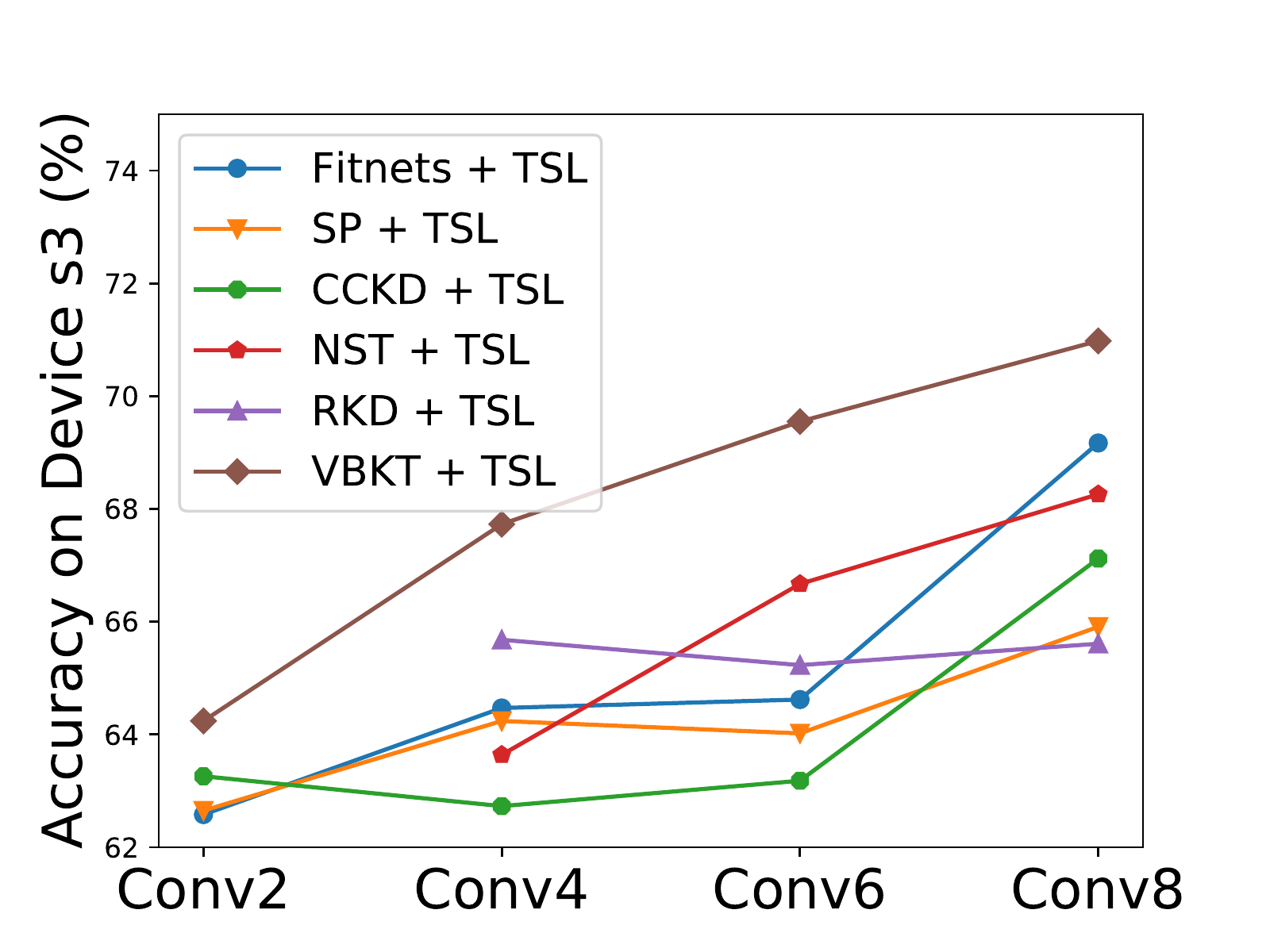}
         \caption{}
         \label{fig:res_layers_d3}
     \end{subfigure}
     \hspace{-0.5cm}
     \begin{subfigure}[b]{0.25\textwidth}
         \centering
         \includegraphics[width=\textwidth]{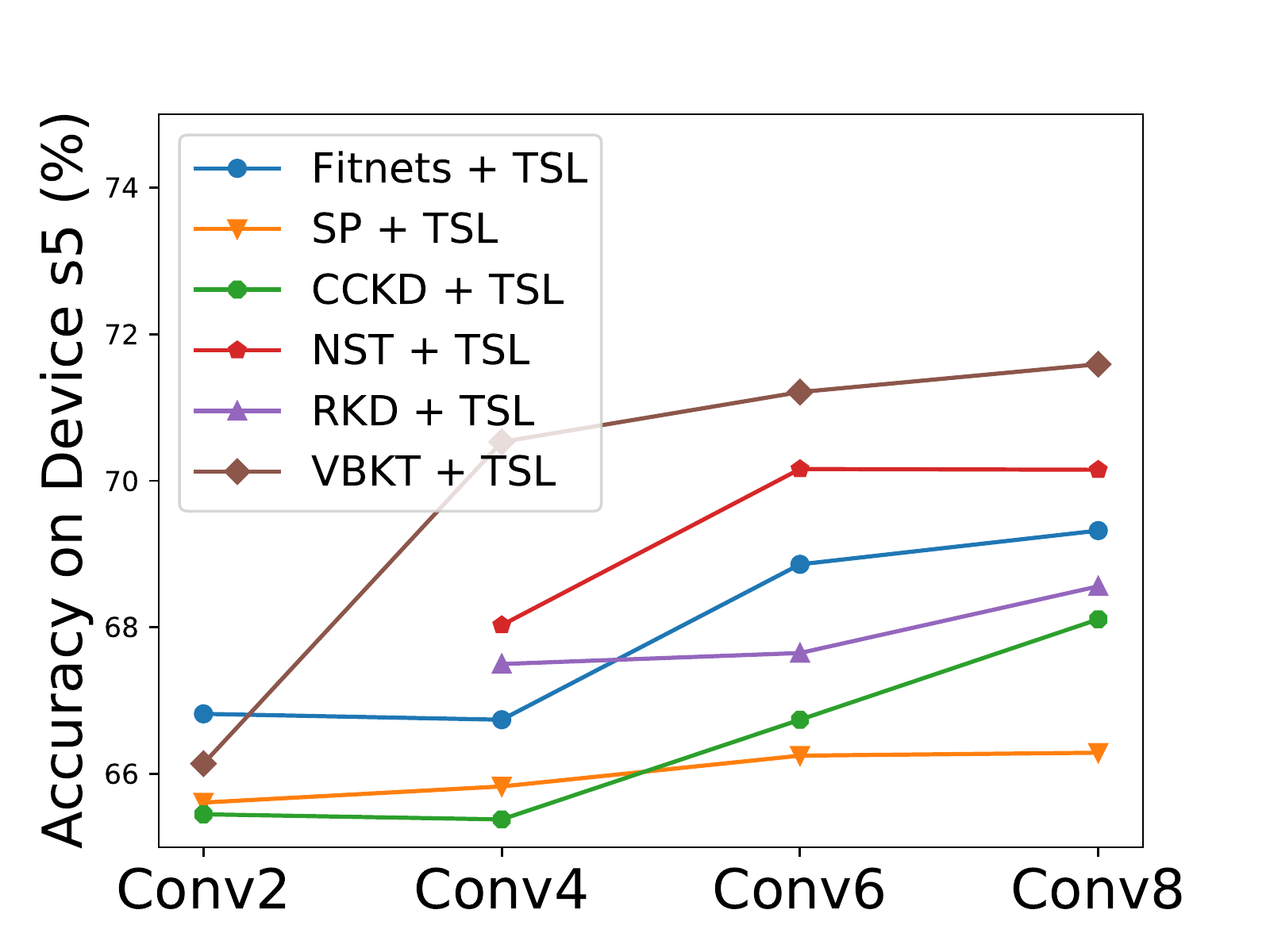}
         \caption{}
         \label{fig:res_layers_d5}
     \end{subfigure}
\vspace{-0.6cm}
        \caption{Evaluation results for using different hidden layers of FCNN model, on two target devices: (a) Device s3 and (b) Device s5. The basic TSL is combined with all methods.}
        \label{fig:res_layers}
\vspace{-0.5cm}
\end{figure}

\vspace{-0.2cm}
\subsection{Effects of Hidden Embedding Depth}
\vspace{-0.1cm}
In our basic setup, the hidden embedding before the last convolutional layer is utilized for modeling latent variables. Ablation experiments are further carried out to investigate the effects of using different-layer hidden embedding. Experiments are performed on FCNN since it has a sequential architecture with stacked convolutional layers, namely 8 convolutional layers and 1$\times$1 convolutional layer for outputs. Results are shown in Figure~\ref{fig:res_layers}. Methods use all hidden layers, like COFD, are not covered here. We don't have results for RKD and NST on Conv2 due to they exceeds the available memory on our machine. From the results we can see that the best performance is obtained from last layer for most of the assessed approaches. Moreover, the hidden embedding closer to the model output allows for a higher accuracy than that closer to the input. Therefore we can argue that late features are better than early features in transferring knowledge across domains. That is in line with what is observed in \cite{romero2014fitnets, huang2017like}. Finally, VBKT attains a very competitive ASC accuracy and consistently outperforms other methods independently of the selected hidden layers.

\vspace{-0.2cm}
\subsection{Visualization of Intra-class Discrepancy}
\begin{figure}[t]
     \centering
     \vspace{-0.2cm}
     \begin{subfigure}[b]{0.17\textwidth}
         \centering
         \includegraphics[width=\textwidth]{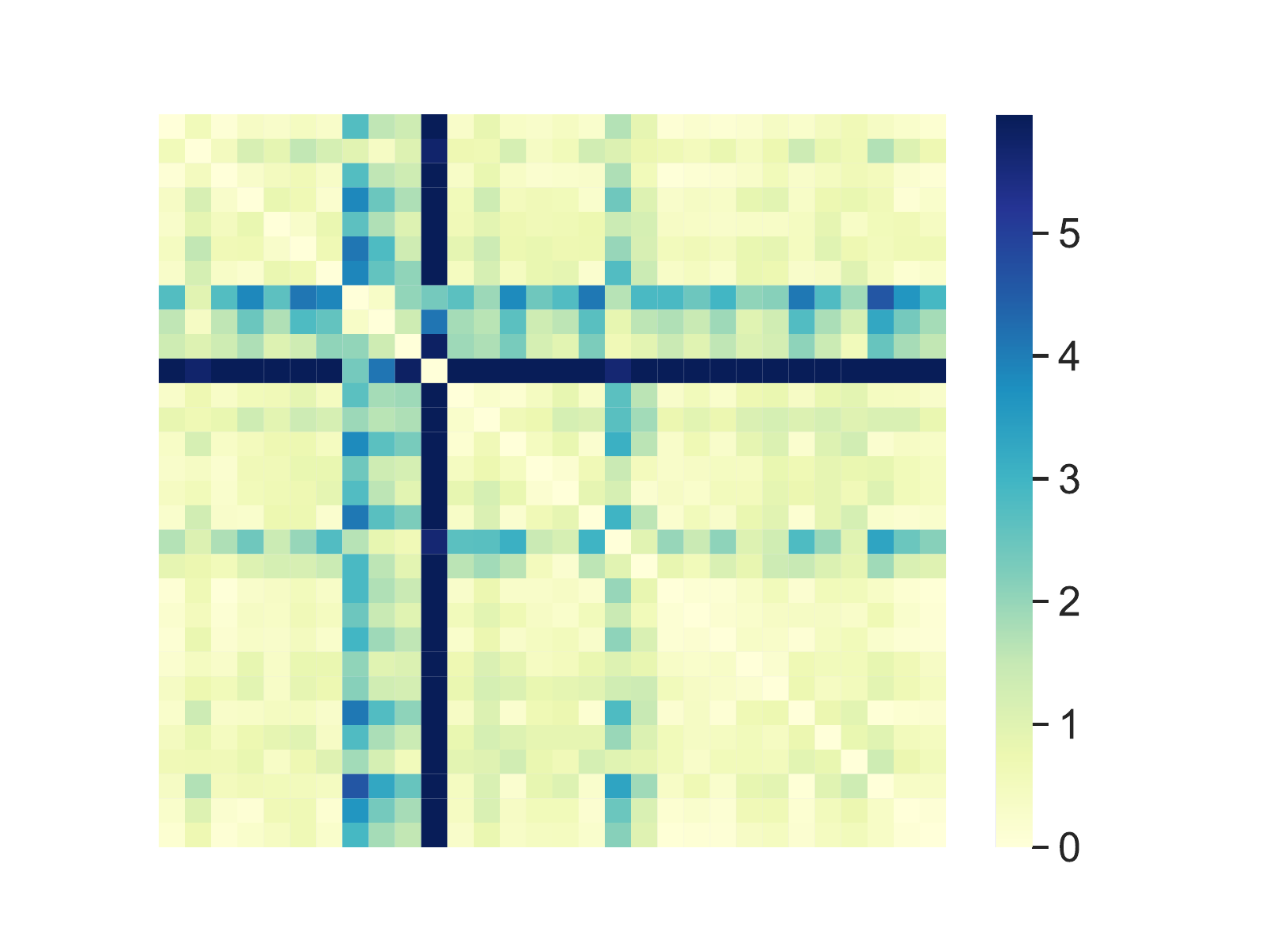}
         \vspace{-0.65cm}
         \caption{TSL}
         \vspace{-0.1cm}
         \label{fig:intra_kd}
     \end{subfigure}
     \hspace{-0.5cm}
     \begin{subfigure}[b]{0.17\textwidth}
         \centering
         \includegraphics[width=\textwidth]{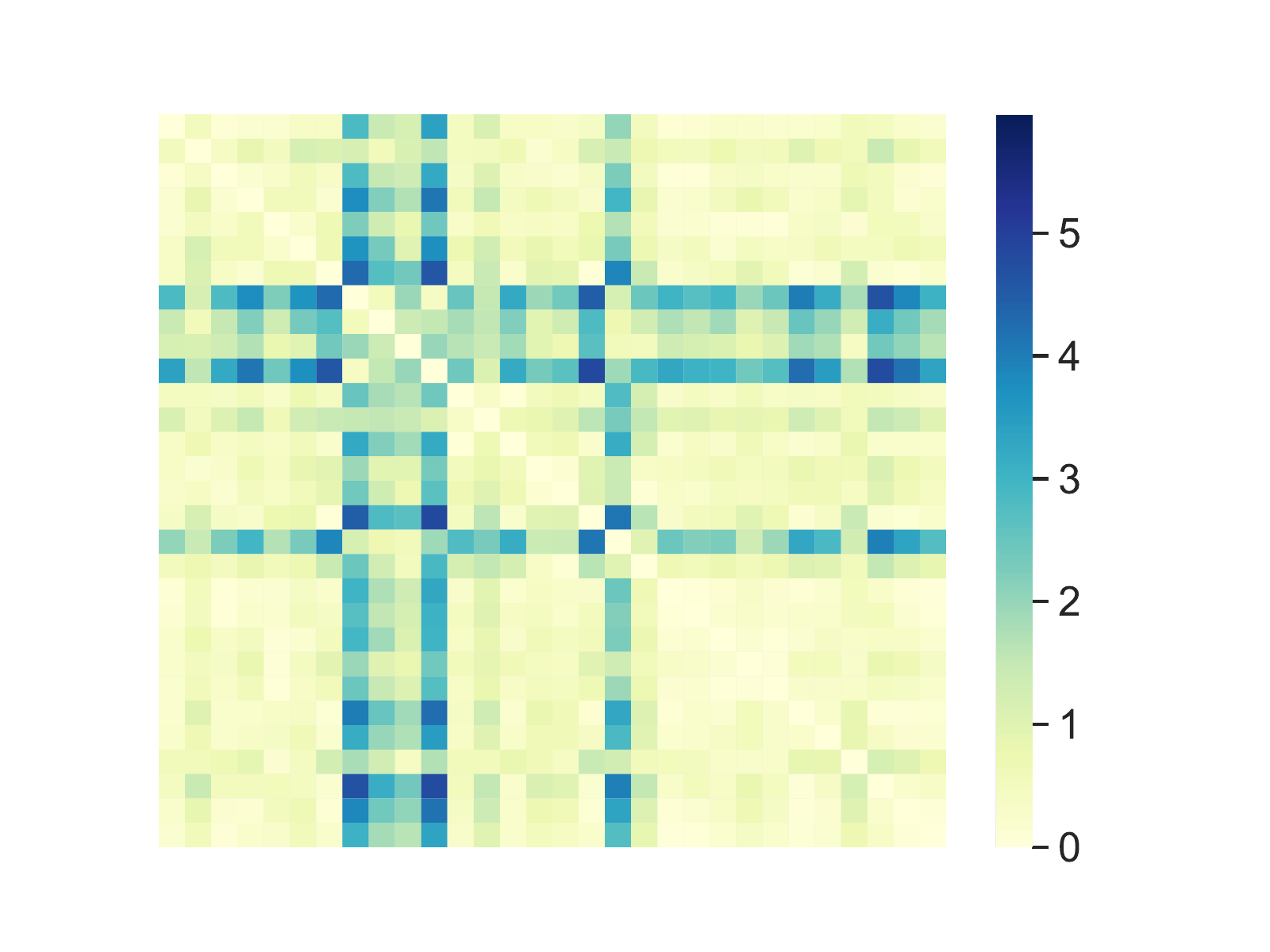}
         \vspace{-0.65cm}
         \caption{Fitnets}
         \vspace{-0.1cm}
         \label{fig:intra_fitnets}
     \end{subfigure}
     \hspace{-0.6cm}
     \begin{subfigure}[b]{0.17\textwidth}
         \centering
         \includegraphics[width=\textwidth]{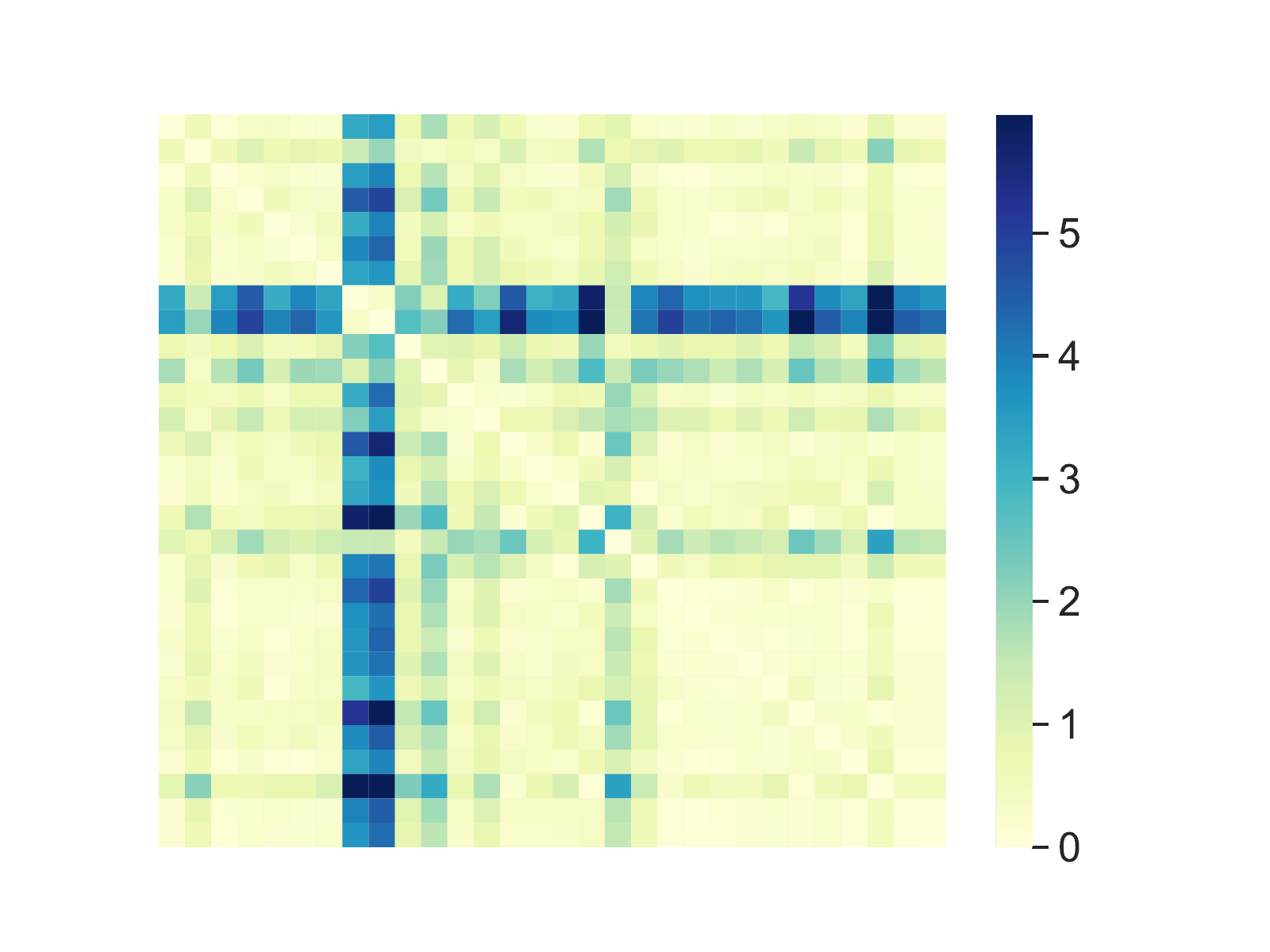}
         \vspace{-0.65cm}
         \caption{AT}
         \vspace{-0.1cm}
         \label{fig:intra_at}
     \end{subfigure}

     \begin{subfigure}[b]{0.17\textwidth}
         \centering
         \includegraphics[width=\textwidth]{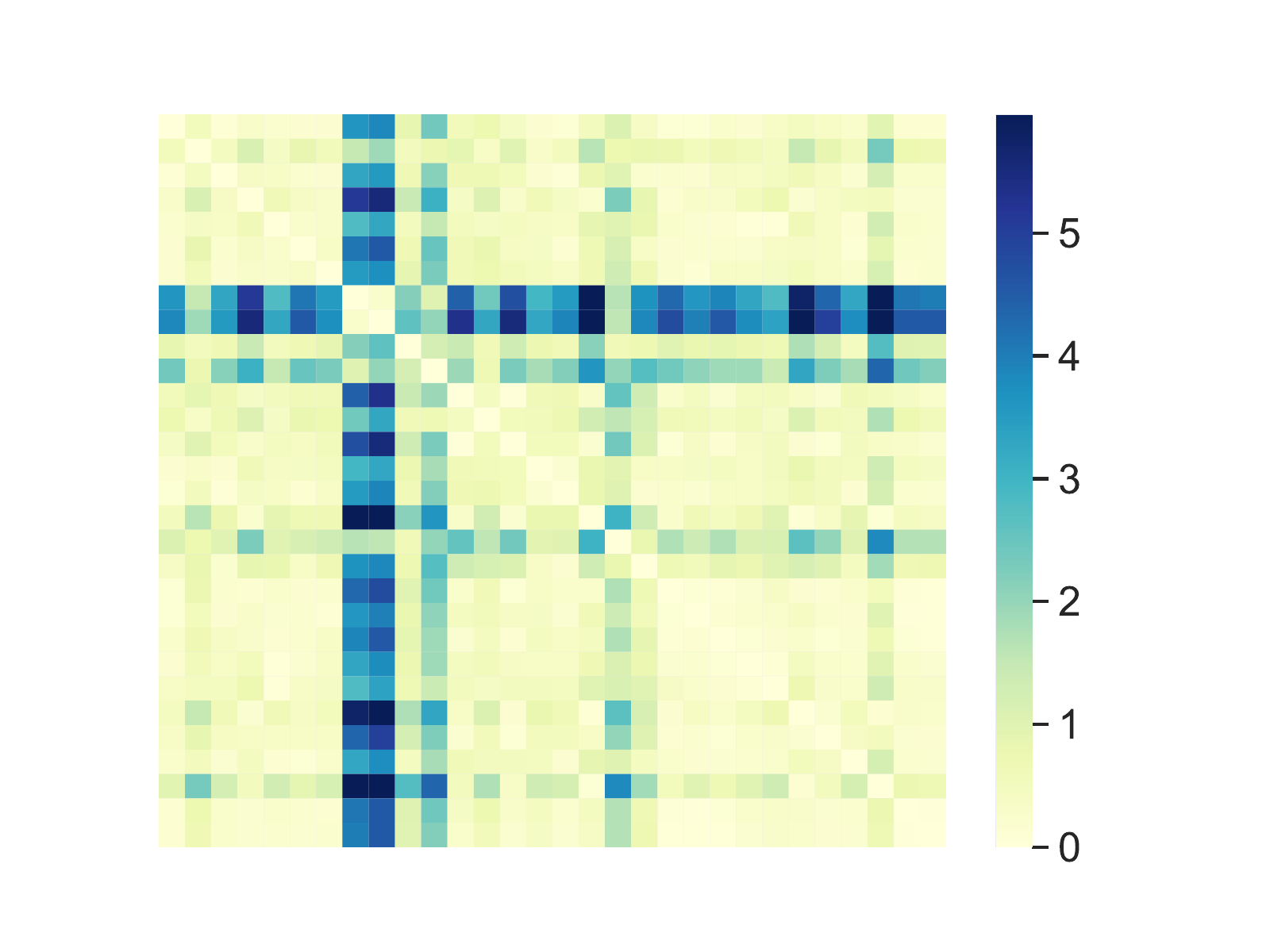}
         \vspace{-0.65cm}
         \caption{SP}
         \vspace{-0.1cm}
         \label{fig:intra_sp}
     \end{subfigure}
     \hspace{-0.6cm}
     \begin{subfigure}[b]{0.17\textwidth}
         \centering
         \includegraphics[width=\textwidth]{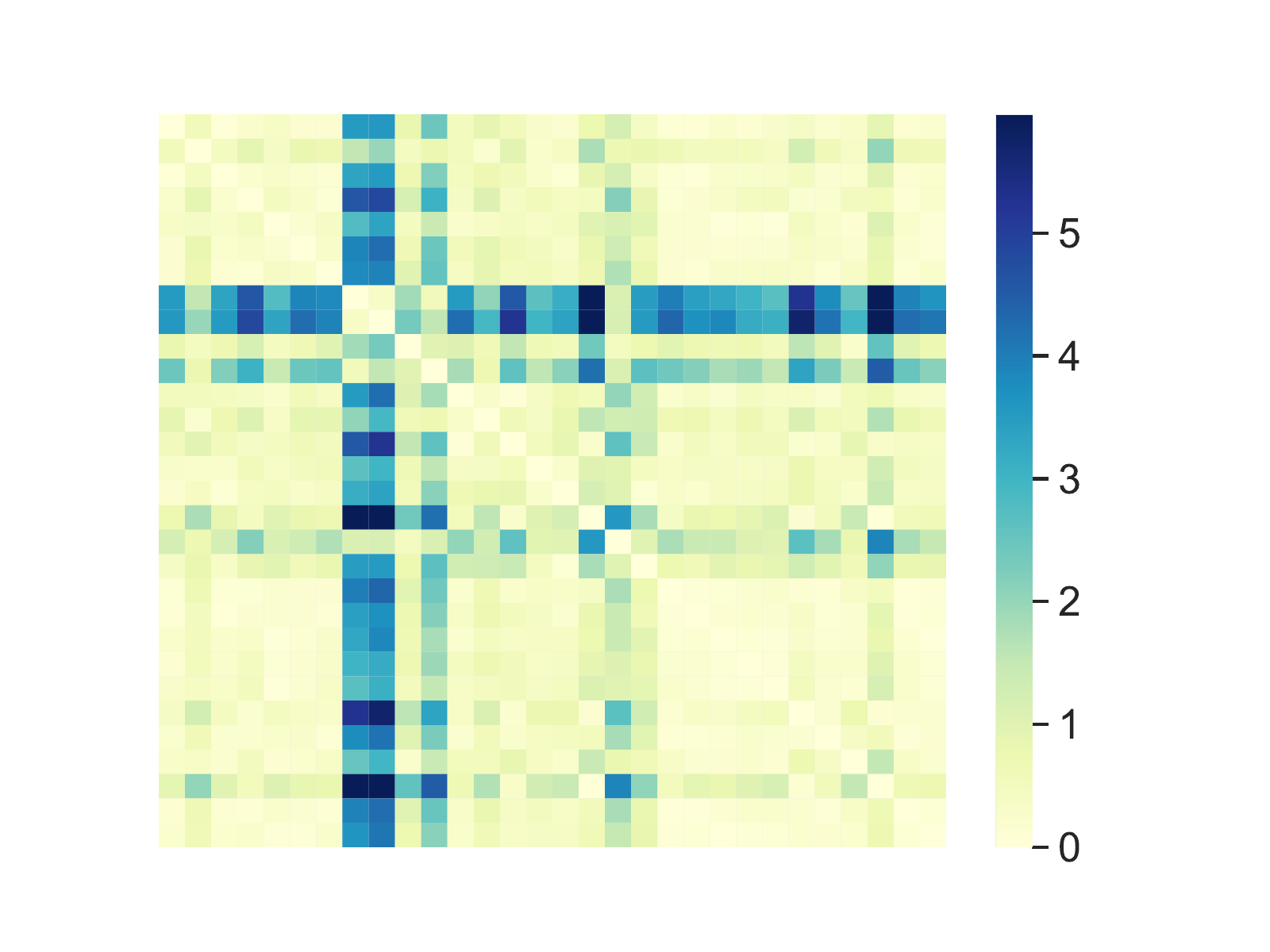}
         \vspace{-0.65cm}
         \caption{CCKD}
         \vspace{-0.1cm}
         \label{fig:intra_cckd}
     \end{subfigure}
     \hspace{-0.6cm}
     \begin{subfigure}[b]{0.17\textwidth}
         \centering
         \includegraphics[width=\textwidth]{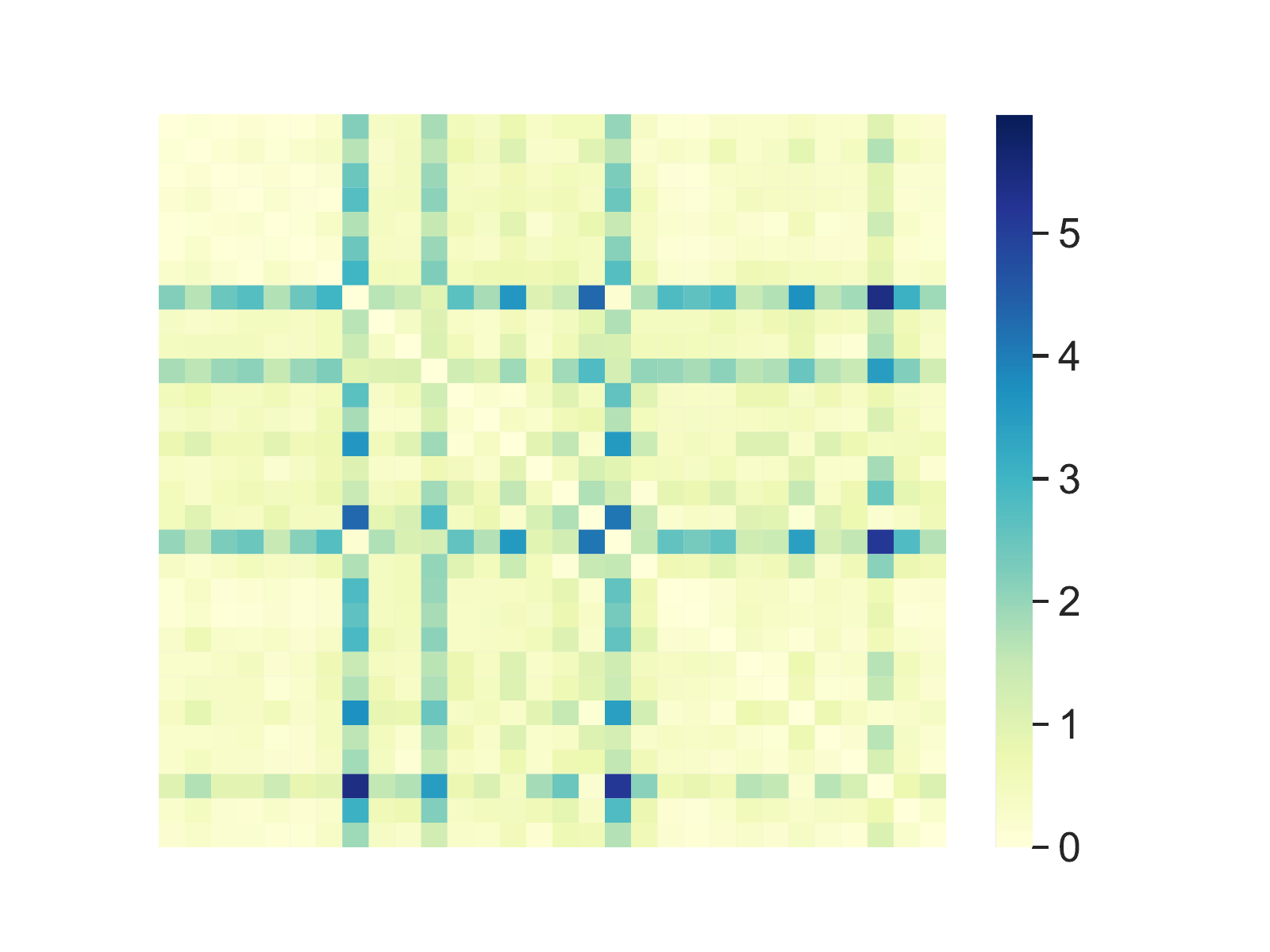}
         \vspace{-0.65cm}
         \caption{VBKT}
         \vspace{-0.1cm}
         \label{fig:intra_vbkt}
     \end{subfigure}
\vspace{-0.1cm}
    \caption{Visualized heatmaps of the intra-class discrepancy between target outputs. FCNN on target device s5 is used. }
    \label{fig:hetmap_intra}
\vspace{-0.5cm}
\end{figure}

\vspace{-0.1cm}
To better understand the effectiveness of the proposed VBKT approach, we compare the intra-class discrepancy between target model outputs (before softmax). The visualized heatmap results are shown in (a)-(f) of Figure~\ref{fig:hetmap_intra}. Here we randomly select 30 samples from the same class and compute $L_2$ distance between model outputs of each two. Thus each cell in subnets of Figure~\ref{fig:hetmap_intra} represents the discrepancy between two outputs, as the darker color means bigger intra-class discrepancy. From these visualization results we can argue that the one obtained by our proposed VBKT approach in Figure~\ref{fig:intra_vbkt} has consistently smaller intra-class discrepancy than those produced by others, implying that VBKT brings up more discriminative information and results in a better cohesion of instances from the same class.

\vspace{-0.2cm}
\section{Conclusion}
\label{sec:con}
\vspace{-0.1cm}
In this study, we propose a variational Bayesian approach to address the cross-domain knowledge transfer issues when deep models are used. Different from previous solutions, we propose to transfer knowledge via prior distributions of deep latent variables from the source domain. We cast the problem into learning distributions of latent variables in deep neural networks. In contrast to conventional maximum a posteriori estimation, a variational Bayesian inference algorithm is then formulated to approximate the posterior distribution in the target domains. We assess the effectiveness of our proposed VB approach on the device adaptation tasks for the DCASE2020 ASC data set. 
Experimental evidence clearly demonstrate that the target model obtained with our proposed approach outperforms all other tested methods in all tested conditions.

\clearpage

\begin{spacing}{0.85}
\footnotesize
\bibliographystyle{IEEEbib}
\bibliography{refs}

\end{spacing}
\end{document}